\documentclass[10pt,conference]{IEEEtran}
\usepackage{graphicx} \graphicspath{ {Figures/} }
\usepackage{blindtext}
\usepackage{comment}
\usepackage[bookmarks,hidelinks,pdfpagelabels=false]{hyperref}

\usepackage{booktabs} % For formal tables

\hypersetup{
    pdftitle={Reasoning about Domain Specifics Languages in Model-driven Architecture},
    pdfauthor={Bedir Tekinerdogan, Ferhat Erata},
    pdfcreator={Ferhat Erata},
    pdfkeywords={Model-driven Engineering, Formal Specification, Domain Specific Modeling, Reasoning, Semantics},
    pdfsubject={International Conference on Software Architecture},
    bookmarksnumbered=true,
    pdfstartview=Fit,
    bookmarksnumbered=true,
    bookmarksopenlevel=\maxdimen,
}

\usepackage{ifpdf}
\newcommand{\T}{Tarski}
\newcommand{\MW}{ModelWriter}
\newcommand{\E}{SIDP}
\newcommand{\case}{System Installation Design Principles (SIDP)}

\newcommand{\sem}[1]{\texttt{\footnotesize{#1}}}
\newcommand{\nl}[1]{\textit{#1}}

\usepackage{blindtext}
\usepackage{amsmath}
\usepackage{amssymb}
\usepackage{mathtools}
\usepackage[frozencache=true,cachedir=minted-cache]{minted} 
\usepackage{nameref}
\usepackage{enumitem}
\usepackage{tikz}
\usetikzlibrary{calc}
\usetikzlibrary{positioning}
\usetikzlibrary{shapes.geometric}
\usetikzlibrary{trees}
\usetikzlibrary{arrows}
\usetikzlibrary{graphs}
\usetikzlibrary{shapes.geometric}
\usetikzlibrary{shapes.misc}
\tikzstyle{block}=[draw opacity=0.7,line width=1.0cm]

\tikzset{
modal/.style={>=stealth',shorten >=1pt,shorten <=1pt,auto,node distance=1.5cm, semithick},
world/.style={circle,draw,minimum size=0.5cm,fill=gray!15},
box/.style={rectangle,draw,fill=gray!15, minimum size=0.5cm},
triangle/.style={regular polygon, regular polygon sides=3,draw,minimum size=0.7cm,fill=gray!15},
ghost/.style={circle,draw,minimum size=0.5cm,fill=white,draw=white, fill opacity=0},
atom/.style={rectangle,minimum size=0.5cm,fill=green!50!black!20,draw=green!50!black},
structure/.style={=>stealth’,shorten >=1pt,shorten <=1pt,auto,node distance=25mm, thick, minimum size=5mm},
set/.style={rectangle, minimum size=0.5cm,fill=green!50!black!20,draw=green!50!black, font=\sffamily\scriptsize},
lattice/.style={=>stealth’,shorten >=1pt,shorten <=1pt, auto, node distance=1cm, semithick, minimum size=3mm},
every node/.style={font=\sffamily\footnotesize},
point/.style={circle,draw,inner sep=0.5mm,fill=black},
scriptsize/.style={font=\sffamily\scriptsize},
tiny/.style={font=\sffamily\tiny},
highlight/.style={fill=red!50!black!20,draw=red!50!black},
comment/.style={rectangle, fill=white,text opacity=1,fill opacity=0, draw opacity=0, minimum size=0.5cm,font=\itshape\sffamily\tiny},
reflexive above/.style={->,loop,looseness=7,in=120,out=60},
reflexive below/.style={->,loop,looseness=7,in=240,out=300},
reflexive left/.style={->,loop,looseness=7,in=150,out=210},
reflexive right/.style={->,loop,looseness=7,in=30,out=330}
}

\DeclareMathOperator{\rsquare}{\mathrel{\square}}

\DeclareMathOperator{\requires}{\mathrel{requires}}
\DeclareMathOperator{\require}{\mathrel{require}}
\DeclareMathOperator{\refines}{\mathrel{refines}}

\DeclareMathOperator{\contains}{\mathrel{contains}}
\DeclareMathOperator{\conflicts}{\mathrel{conflicts}}

\DeclareMathOperator{\equals}{\mathrel{equals}}

\usepackage{color}

\usepackage{cite}
\usepackage{amsmath}
\usepackage{amssymb}
\usepackage{algorithmic}
\usepackage{array}
\ifCLASSOPTIONcompsoc
  \usepackage[caption=false,font=normalsize,labelfont=sf,textfont=sf]{subfig}
\else
  \usepackage[caption=false,font=footnotesize]{subfig}
\fi
\usepackage{url}
\begin{document}
\title{ModelWriter: Text \& Model-Synchronized Document Engineering Platform}



\author{
    \IEEEauthorblockN{
        \href{mailto:ferhat@ieee.org}{Ferhat Erata}\IEEEauthorrefmark{1}\IEEEauthorrefmark{4},
        \href{mailto:claire.gardent@loria.fr}{Claire Gardent}\IEEEauthorrefmark{2},
        \href{mailto:bikash.gyawali@loria.fr}{Bikash Gyawali}\IEEEauthorrefmark{2},
        \href{mailto:anastasia.shimorina@loria.fr}{Anastasia Shimorina}\IEEEauthorrefmark{2},
        \href{mailto:yvan.lussaud@obeo.fr}{Yvan Lussaud}\IEEEauthorrefmark{6}, 
        \href{mailto:bedir.tekinerdogan@wur.nl}{Bedir Tekinerdogan}\IEEEauthorrefmark{1}, \\
        \href{mailto:geylani.kardas@ege.edu.tr}{Geylani Kardas}\IEEEauthorrefmark{5}\IEEEauthorrefmark{7} and 
        \href{mailto:anne.monceaux@airbus.com}{Anne Monceaux}\IEEEauthorrefmark{3}}

\IEEEauthorblockA{
    \IEEEauthorrefmark{1}Information Technology Group, Wageningen University and Research Centre, The Netherlands}
\IEEEauthorblockA{
    \IEEEauthorrefmark{2}CNRS, LORIA, UMR 7503 Vandoeuvre-les-Nancy, F-54500, Nancy, France}
\IEEEauthorblockA{
    \IEEEauthorrefmark{3}System Engineering Platforms, Airbus Group Innovations, Toulouse, France}
\IEEEauthorblockA{
    \IEEEauthorrefmark{4}UNIT Information Technologies R\&D Ltd., Izmir, Turkey}
\IEEEauthorblockA{
    \IEEEauthorrefmark{5}Ege University, International Computer Institute, Izmir, Turkey}
\IEEEauthorblockA{
    \IEEEauthorrefmark{7}Ko\c{c}Sistem Information and Communication Services Inc. Istanbul, Turkey}
\IEEEauthorblockA{
    \IEEEauthorrefmark{6}OBEO, Nantes, France}
}    
%Email: ferhat@computer.org


\maketitle

\begin{abstract}
The ModelWriter platform provides a generic framework for automated traceability analysis. In this paper, we demonstrate how this framework can be used to trace the consistency and completeness of technical documents that consist of a set of System Installation Design Principles used by Airbus to ensure the correctness of aircraft system installation. We show in particular, how the platform allows the integration of two types of reasoning: reasoning about the meaning of text using semantic parsing and description logic theorem proving; and reasoning about document structure using first-order relational logic and finite model finding for traceability analysis.
\begin{center}
\fbox{\bf\small{\url{https://itea3.org/project/modelwriter.html}}} %https://modelwriter.github.io/semanticparser/screencast/
\end{center}
\end{abstract}

% no keywords https://youtu.be/Ba4yqoEFaSs

% \fi
\IEEEpeerreviewmaketitle

% !TEX root =  Main.tex
\section{Introduction} 
\label{sec:intro}

% (should ne ver be an issue)

The complexity of software systems in safety critical domains (e.g. avionics and automotive) has significantly increased over the years. Development of such systems requires various phases which result in several artifacts (e.g., requirements documents, architecture models and test cases). In this context, traceability~\cite{Ramesh:2001aa, SWEBOK:IEEEComputerSociety:2014} not only establishes and maintains consistency between these artifacts but also helps guarantee that each requirement is fulfilled by the source code and test cases properly cover all requirements, a very important objective in safety critical systems and the standards they need to comply with DO-178C (Software Considerations in Airborne Systems and Equipment Certification)~\cite{DO178C} and ISO-26262 (Road Vehicles - Functional Safety)~\cite{ISO26262}. As a result, the engineers have to establish and maintain several types of traces, having different semantics, between and within various development artifacts. 

Traceability is a quality concern that helps users understand each and every steps in the development or even the end to end life cycle of a product. Its implementation is highly contextual as the key artifacts produced or used along a process differ depending on the product. We want to provide a framework for users to specify which artifacts they want to precisely identify and monitor and what is the meaning for trace links between these artifacts.

The considered Artifacts represented in our context by trace locations might be of different levels of granularity, ranging from a complete document or model to fragments of text or code. Focusing on documents and text, both the structure and the content might be used to reason about traceability.

To this end, \MW~platform provides a generic traceability analysis applicable to Text \& Model artifacts. Trace locations can be fragments of text, elements of an architectural model, and parts of program codes. Traces are relations between trace locations. \MW~platform allows axiomatization of these relations and reasoning about them, i.e. supporting traceability analysis for different types of artifacts. 

In this paper, we focus on demonstrating the features of \MW~platform for the traceability analysis applied to technical documentation. A particular challenge in this use case is to take into account the meaning of natural language. We integrate techniques from Natural Language Processing (NLP) and Automated Reasoning to reason both about the meaning and about the structure of text. We use techniques from semantic parsing to assign formal meaning representations to NL text. We then use techniques from theorem proving and model building to infer traceability relations between text fragments (here SIDPs), to check consistency and to ensure completeness. 

%\T~is developed for environments, requiring maintenance of various artifacts, within the context of our research~\cite{ModelWriter, Assume} in collaboration with Ford-Otosan~\cite{FordOtosan}, Airbus~\cite{Airbus} and Havelsan~\cite{Havelsan}.

%However, these are listed in documents and disconnected from formal models. Being able to translate text to model would permit to check consistency.

% !TEX root =  Main.tex
\section{The Airbus SIDP Usecase} 
\label{sec:usecase}

We illustrate the workings of the \MW~platform based on a set of System Installation Design Principles (SIDP) used by Airbus to ensure the correctness of aircraft design. A SIDP rule is actually a kind of system installation requirement, that is a description of system properties which should be fulfilled. 
In this usecase, SIDPs are trace locations and there are five types of trace links defined between trace locations, namely \textsc{contains}, \textsc{refines}, \textsc{conflicts}, \textsc{equals}, and \textsc{requires}. In the following, we informally give the meaning of the trace-types.

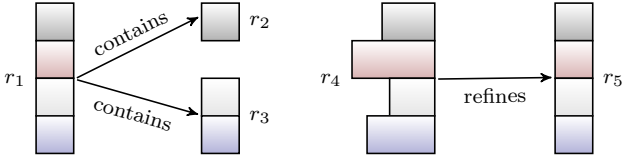
\begin{figure}[H]
\centering
    \begin{tikzpicture}[modal,node distance=2.2cm]
        \node (A) [box, top color=white, bottom color=gray!50!black!40] {};
        \node (B) [box, below= -\pgflinewidth of A, top color=white, bottom color=red!50!black!30] [label={[yshift=2mm]south west:$r_1$}] {};
        \node (C) [box, below= -\pgflinewidth of B, top color=white, bottom color=white!50!black!20] {};
        \node (D) [box, below= -\pgflinewidth of C, top color=white, bottom color=blue!50!black!30] {};
        
        \node (E) [box, right of=A, top color=white, bottom color=gray!50!black!40] [label={right:$r_2$}]  {};
        \node (F) [box, right of=C, top color=white, bottom color=white!50!black!20] [label={[yshift=2mm]south east:$r_3$}]  {};
        \node (G) [box, below= -\pgflinewidth of F, top color=white, bottom color=blue!50!black!30] {};

        \path [->]
            (B.south east) edge node[sloped, anchor=south] {$\contains$} (E.mid west)
            (B.south east) edge node[sloped, anchor=north] {$\contains$} (G.north west);
            
        \node (A_c) [box, node distance=2.5cm, right of=E, top color=white, bottom color=gray!50!black!40, minimum width=0.7cm] {};
        \node (B_c) [box, node distance=0.5cm, below=of A_c.east, anchor=east, top color=white, bottom color=red!50!black!30, minimum width=1.1cm] [label={[yshift=2mm]south west:$r_4$}] {};
        \node (C_c) [box, node distance=0.5cm, below=of B_c.east, anchor=east, top color=white, bottom color=white!50!black!20, minimum width=0.6cm] {};
        \node (D_c) [box, node distance=0.5cm, below=of C_c.east, anchor=east, top color=white, bottom color=blue!50!black!30, minimum width=0.9cm] {};
        
        \node (E_c) [box, right of=A_c, top color=white, bottom color=gray!50!black!40] {};
        \node (F_c) [box, below= -\pgflinewidth of E_c, top color=white, bottom color=red!50!black!30] [label={[yshift=2mm]south east:$r_5$}]{};
        \node (G_c) [box, below= -\pgflinewidth of F_c, top color=white, bottom color=white!50!black!20] {};
        \node (H_c) [box, below= -\pgflinewidth of G_c, top color=white, bottom color=blue!50!black!30] {};

        \path [->]
            (B_c.south east) edge node[sloped, anchor=north] {$\refines$} (G_c.north west);
            
    \end{tikzpicture}
\vspace{-0.5em}
\caption{$r_1$ \textit{contains} $r_2$ and $r_3$, $r_4$ \textit{refines} $r_5$ }
\label{fig:informal_contains_refines}
\vspace{-0.5em}
\end{figure}

\textit{Rule} $r_1$ \textit{contains} \textit{Rule} $r_2 \dots r_n$ if $r_2 \dots r_n$ are parts of the whole $r_1$ (part-whole hierarchy). The contained rule is a sub-rule of containing rule. \textit{Rule} $r_1$ \textit{refines} another \textit{Rule} $r_2$ if $r_1$ is derived from $r_2$ by adding more details to its properties. The refined rule can be seen as an abstraction of the detailed rules. In Fig.~\ref{fig:informal_contains_refines} \textit{contains} and \textit{refines} traces are illustrated. Each box represents a property of the corresponding rule.

\textit{Rule} $r_1$ \textit{conflicts} with \textit{Rule} $r_2$ if the fulfillment of $r_1$ excludes the fulfillment of $r_2$ and vice versa. The existence of a conflict trace indicates an inconsistency between two rules. \textit{Rule} $r_1$ \textit{equals} to \textit{Rule} $r_2$ if $r_1$ states exactly the same properties with their constraints with $r_2$ and vice versa. \textit{Rule} $r_1$ \textit{requires} \textit{Rule} $r_2$ if $r_1$ is fulfilled only when $r_2$ is fulfilled. The required rule can be seen as a pre-condition for the requiring rule. In the following Fig.~\ref{fig:informal_conflicts_equals_requires} \textit{conflicts}, \textit{equals} and \textit{requires} traces are illustrated. 

\begin{figure}[H]
\vspace{-0.5em}
\centering
    \begin{tikzpicture}[modal,node distance=2.2cm]
        %conflicts
        \node (A) [box, top color=white, bottom color=gray!50!black!40] [label={left:$r_1$}]{};
        \node (E) [box, right of=A, top color=white, bottom color=red!50!black!30] [label={right:$r_2$}]  {};

        \path [<->]
            (A) edge node[sloped, anchor=south] {$\conflicts$} (E);
            
        %equals
        \node (Ae) [box, node distance=2.2cm, right of=E, top color=white, bottom color=gray!50!black!40] [label={left:$r_1$}]{};
        \node (Ee) [box, right of=Ae, top color=white, bottom color=gray!50!black!40] [label={right:$r_2$}]  {};

        \path [<->]
            (Ae) edge node[sloped, anchor=south] {$\equals$} (Ee);
            
        %requires
        \node (Ar) [box, node distance=1.0cm, below of=E, top color=white, bottom color=gray!50!black!40] [label={left:$r_1$}]{};
        \node (Er) [box, right of=Ar, top color=white, bottom color=red!50!black!30] [label={right:$r_2$}]  {};

        \path [->]
            (Ar) edge node[sloped, anchor=south] {$\requires$} (Er);
    \end{tikzpicture}
\vspace{-0.5em}
\caption{Illustration of ``\textit{conflicts, equals and requires}''}
\label{fig:informal_conflicts_equals_requires}
\vspace{-0.5em}
\end{figure}
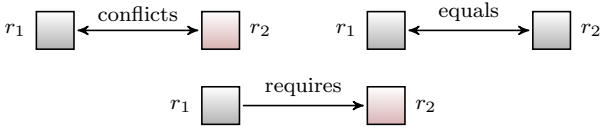

Given a set of SIDPs, the \MW~platform can be used to check completeness and consistency as follows. First, SIDPs are parsed and assigned Description Logic formulae representing their meaning (cf. Section~\ref{subsec:semanticparsing}). Second, traces are either manually specified by the end user or can be inferred using semantic parsing and DL theorem proving (cf. Section~\ref{subsec:dltp}). Third, new traces can be inferred upon existing ones using Relational Logic (cf. Section~\ref{subsec:formalization}) and Model Finding (cf. Section~\ref{subsec:inferrence_traces}). Importantly, the inference of trace links allows for the detection of missing or inconsistent SIDPs.

Table~\ref{tab:sidps} illustrates this process. Given the SIDPs $r_1$-$r_6$, \textsc{conflicts} and \textsc{refines} trace links are first inferred using semantic parsing and the Hermit theorem prover~\cite{hermit} (DL lines in the table). 

\begin{table}[ht]
\vspace{-1.0em}
\caption{Example SIDPs and inference of Trace Links}\label{tab:sidps}
\vspace{-1.0em}
\begin{tabular}{ll}
 \toprule
 Nr. & Artifact Annotations (Trace-locations)\\ 
 \midrule
 $\text{r}_\text{1}$ &Bracket shall be used in hydraulic area Alpha \\
 $\text{r}_\text{2}$ &Adhesive bonded bracket shall be used in hydraulic area  \\
 $\text{r}_\text{3}$ &Adhesive bonded bracket shall be used in hydraulic area Alpha \\
 $\text{r}_\text{4}$ &Bracket shall be used in hydraulic area \\
 $\text{r}_\text{5}$ &Bracket shall be installed in hydraulic area \\
 $\text{r}_\text{6}$ &Bracket shall be installed in fuel tank\\
 \bottomrule
\end{tabular}

\vspace{0.5em}

\begin{tabular}{l l @{\hspace{5.0em}} l l @{\hspace{3.0em}}}
% \toprule
 Nr. & Inferred Traces & Nr. & Inferred Traces\\ 
 \midrule
 $\text{DL}_\text{1}$  &$conflicts(r_5,r_6)$ &$\text{RL}_\text{1}$  &$conflicts(r_6,r_4)$ \\
 $\text{DL}_\text{2}$  &$refines(r_3,r_2)$   &$\text{RL}_\text{2}$  &$requires(r_1,r_5)$ \\
 $\text{DL}_\text{3}$  &$refines(r_2,r_4)$   &$\text{RL}_\text{3}$  &$conflicts(r_6,r_1)$ \\
 $\text{DL}_\text{4}$  &$refines(r_1,r_4)$   &$\text{RL}_\text{4}$  &$requires(r_2,r_5)$ \\
 $\text{DL}_\text{5}$  &$requires(r_4,r_5)$  &$\text{RL}_\text{5}$  &$conflicts(r_2,r_6)$ \\
 \bottomrule
\end{tabular}
\vspace{-0.5em}
\end{table}

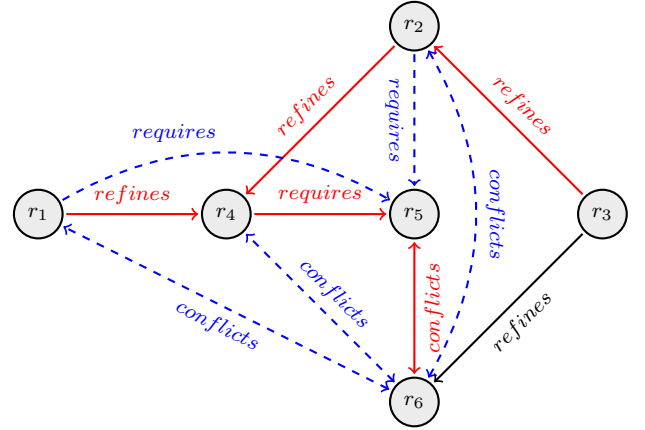
\begin{figure}[ht]
\begin{center}
\vspace{-1.0em}
    \begin{tikzpicture}[structure]
        \node (G0) [ghost] {$r_2$};
        \node (G1) [ghost, right of=G0] {$r_2$};
        \node (2) [world, right of=G1] {$r_2$};
        \node (4) [world, below of=G1] {$r_4$};
        \node (1) [world, left of=4] {$r_1$};
        \node (5) [world, right of=4] {$r_5$};
        \node (3) [world, right of=5] {$r_3$}; %fill=red!50!black!20,draw=red!50!black
        \node (6) [world, below of=5] {$r_6$}; %fill=red!50!black!20,draw=red!50!black

        \path [->][red, sloped, anchor=south]
            (1) edge node {$refines$} (4);
                
        \path [->][red, sloped, anchor=south]
            (3) edge node {$refines$} (2);
                
        \path [->][red, sloped, anchor=south]
            (2) edge node {$refines$} (4);
                
        \path [->][red, sloped, anchor=south]
            (4) edge node {$requires$} (5);
            
        \path [->][sloped, anchor=north]
            (3) edge node {$refines$} (6);

        \path [<->] [red]
            (6) edge node[sloped,anchor=north] {$conflicts$} (5);
        
        \path [<->] [blue, dashed]
            (6) edge node[sloped,anchor=south] {$conflicts$} (4);
                
        \path [->] [blue, dashed]
            (2) edge node[sloped,anchor=north] {$requires$} (5);
      
        \path [<->] [blue, dashed]
            (2) edge[bend left] node[sloped,anchor=south] {$conflicts$} (6);

        \path [->] [blue, dashed]
            (1) edge[bend left] node[sloped,above left  ] {$requires$} (5);

        \path [<->] [blue, dashed]
            (6) edge node[sloped,anchor=north] {$conflicts$} (1);
    \end{tikzpicture}
\vspace{-0.5em}
\caption{Inferred Traces (red traces indicate reasoning using DL, blue indicates reasoning using RL, the black one indicates a manual trace)}
\label{fig:inference}
\vspace{-0.5em}
\end{center}
\end{figure}

For example, the DL formulae obtained by parsing sentences $r_5$ and $r_6$ conflict with each other because the underlying ontology to which these axioms are added specifies that concepts ``hydraulic area'' and ``fuel tank'' are disjoint. Similarly, the axiom obtained for the sentence $r_2$ refines the axiom obtained for $r_4$ because the ontology specifies that ``Bracket'' is a sub concept of ``Adhesive bonded bracket''. In Fig. \ref{fig:inference}, Table \ref{tab:sidps} is represented as a digraph model in which the nodes represent trace-locations, i.e. SIDP rules listed in the table and edges represents traces. A red edge specifically corresponds to the trace inferred using semantic parsing and DL theorem proving. The black one is an example trace, $refines(r_3, r_6)$ created by the user manually. 

\begin{figure}[H]
\vspace*{-0.5em}
\centering
      \begin{tikzpicture}[structure, world/.append style={minimum size=0.7cm}]
        \node (A) [world] {$a$};
        \node (B) [world, right of=A] {$b$};
        \node (C) [world, below of=B] {$c$};
        \node (D) [world, right of=B] {$a$};
        \node (E) [world, right of=D] {$b$};
        \node (F) [world, below of=E] {$c$};

        \path [->]
                (A) edge node {$refines$} (B)
                (B) edge node[sloped,anchor=south] {$requires$} (C)
                (D) edge node {$requires$} (E)
                (E) edge node[sloped,anchor=south] {$conflicts$} (F);

        \path [->] [blue, dashed]
            (A) edge node[sloped,anchor=north] {$requires$} (C)
            (D) edge node[sloped,anchor=north] {$conflicts$} (F);
      \end{tikzpicture}
\vspace*{-0.5em}
\caption{Inferring ``\textit{requires}" with ``\textit{refines}" and inferring ``\textit{conflicts}" }
\label{fig:reasoning_requires_conflicts}
\vspace*{-0.5em}
\end{figure}
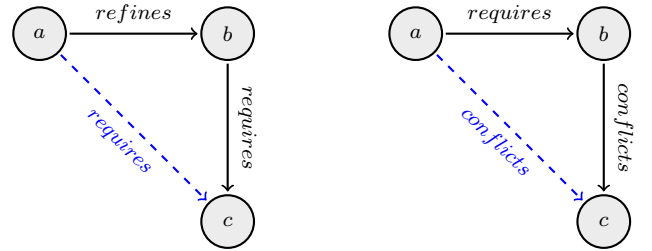

Later, additional trace links are inferred using Relational Model Finding (RL lines in the Table~\ref{tab:sidps} and dashed blue edges on Fig.~\ref{fig:inference}). For instance, as part of the trace semantics of this use case, according to the axiom schema \eqref{axiom:conflict_generation} formalized in Section \ref{subsec:formalization} where \textit{a}, \textit{b} and \textit{c} are artifact elements, if \textit{a} \textit{refines}, \textit{requires} or \textit{contains} \textit{b}, while \textit{b} \textit{conflicts} with \textit{c}, then \textit{a} also \textit{conflicts} with \textit{c}. In this way, \MW~ generates \textsc{conflicts} traces such that combination of $conflicts(r_5,r_6)$ and $requires(r_4,r_5)$ makes $conflicts(r_6,r_4)$; on the other hand, according to axiom schema \eqref{axiom:requires_1} described in Section \ref{subsec:formalization}, the combination of $refines(r_2,r_4)$ and $requires(r_4,r_5)$ generates $requires(r_2,r_5)$ corresponding to the patterns shown in Fig. \ref{fig:reasoning_requires_conflicts}.

Finally, in this example, DL-based reasoning process inferred only one \textsc{conflicts} trace using the meaning of the sentences, i.e. $r_5$ conflicts with $r_6$ whereas the \MW~detects three more \textit{conflicts} traces using the meaning of trace types by means of RL-based reasoning on top of DL-based reasoning. As a result, it can be seen that not only $r_5$ and $r_6$ but also $r_4$, $r_1$, and $r_2$ are inconsistent.

% !TEX root =  Main.tex
\section{Overview of the Approach} 
\label{sec:approach}

We now describe the four main modules making up the \MW~platform. Section~\ref{subsec:semanticparsing} introduces the semantic parser, i.e., the module that converts text to Description Logic formulae. 
Section~\ref{subsec:dltp} explains how the Hermit reasoner can be used to detect \textsc{refines, conflicts} and \textsc{equals} trace links between text fragments (here, SIDPs). Section~\ref{subsec:formalization} shows how Alloy formalism \cite{AlloyBook} can be customized to axiomatize trace types and semantics. Finally, Section~\ref{subsec:inferrence_traces} explains how the KodKod model finder~\cite{KodKodPhdThesis} is used to infer new trace links between SIDPs to detect the inconsistent SIDPs. 

\subsection{Mapping Text to Description Logic Formulae}
\label{subsec:semanticparsing}

The semantic parser used in ModelWriter to convert text to DL formulae is described in details in \cite{gyawali2017mapping}. In what follows, we briefly summarize its working and some evaluation results on a set of 960 SIDPs used for testing. 

The \MW~semantic processing framework combines an automatically derived lexicon, a small hand-written grammar, a parsing algorithm to convert text to DL formulae and a generation algorithm to generate text from DL formulae. This framework is modular, robust and reversible. It is modular in that, different lexicons or grammars may be plugged to meet the requirements of the semantic application being considered. For instance, the lexicon (which relates words and concepts) could be built using a concept extraction tool, i.e. a text mining tool that extracts concepts from text (e.g., \cite{bozsak2002kaon}). And the grammar could be replaced by a grammar describing the syntax of other document styles such as cooking recipes. It is robust in that, in the presence of unknown words, the parser can skip words and deliver a connected (partial) parse. And, it is reversible in that the same grammar and lexicon can be used both for parsing and for generation. Fig.~\ref{fig:parse_gen_process} outlines  our approach showing the interaction of various components.

\tikzstyle{inputText} = [rectangle, rounded corners, minimum width=3cm, minimum height=1cm,text centered, draw=black]
\tikzstyle{regeneratedText} = [rectangle, rounded corners, minimum width=3cm, minimum height=1cm,text centered, draw=black]
\tikzstyle{grammarLexicon} = [rectangle, rounded corners, double=black, double distance =1pt,minimum width=3cm, text centered]
\tikzstyle{general} = [rectangle, rounded corners, minimum width=3cm, minimum height=1cm,text centered, draw=black]
\tikzstyle{smallrect} = [rectangle, rounded corners, text centered, draw=black]
\tikzstyle{processing} = [rectangle, minimum width=3cm, minimum height=1cm,text centered, draw=black, fill=gray!80]

\tikzstyle{arrow} = [thick,->,>=stealth]

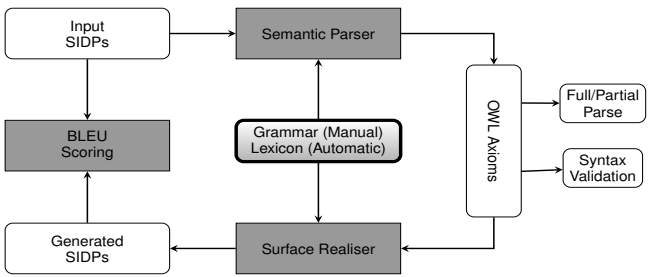
\begin{figure}[H]
  \centering
  \vspace{-0.5em}
  \resizebox{0.48\textwidth}{.15\textheight}
  {
    \begin{tikzpicture}[node distance=2cm]
      \node (InputText) [inputText, align=center] {Input\\SIDPs};
      \node (SemanticParsing) [processing, right=1.2cm of InputText] {Semantic Parser};
      \node (GrammarLexicon) [grammarLexicon, below=1.2cm of SemanticParsing, align=center, top color=white, bottom color=gray!50!black!40, draw] {Grammar (Manual)\\Lexicon (Automatic)}; %The Align option is needed to enable multiline text 
      \node (OWLDL) [general, right=2.2cm of GrammarLexicon, rotate=270,anchor=north] {OWL Axioms};
      \node (SR) [processing, below=1.2cm of GrammarLexicon] {Surface Realiser};
      \node (RegeneratedText) [regeneratedText, left=1.2cm of SR, align=center] {Generated\\SIDPs};
      \node (BLEUScore) [processing, below=1.2cm of InputText, align=center] {BLEU\\Scoring};
      \node (FullORCompleteParse) [smallrect, right = 1.2cm of OWLDL, align=center, yshift=2.25cm] {Full/Partial\\Parse};
      \node (OWLSyntaxValidation) [smallrect, below=0.5cm of FullORCompleteParse, align=center] {Syntax\\Validation};

      \draw [arrow] (InputText) -- (SemanticParsing);
      \draw [arrow] (GrammarLexicon) -- (SemanticParsing);
      \draw [arrow] (GrammarLexicon) -- (SR);
      \draw [arrow] (SemanticParsing) -| (OWLDL);
      \draw [arrow] (OWLDL) |- (SR);
      \draw [arrow] (SR) -- (RegeneratedText);
      \draw [arrow] (RegeneratedText) -- (BLEUScore);
      \draw [arrow] (InputText) -- (BLEUScore);
      \draw [arrow] ([yshift=0.75cm]OWLDL.north) -- (FullORCompleteParse);
      \draw [arrow] ([yshift=-0.6cm,xshift=1cm]OWLDL.south) -- (OWLSyntaxValidation);
    \end{tikzpicture}
  }
  \vspace{-0.5em}
  \caption{Parsing and Generation of Airbus SIDPs.} 
  \label{fig:parse_gen_process} 
  \vspace{-0.5em}
\end{figure}

\begin{figure}[ht]
\vspace{-1.0em}
\begin{center}
\includegraphics[width=0.9\columnwidth]{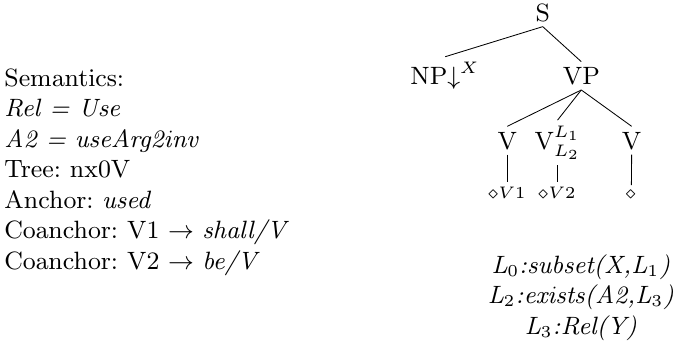}
\caption{Example Lexical Entry and Grammar Unit }
\label{fig:lexicon}
\vspace{-1.0em}
\end{center} 
\vspace{-0.5em}
\end{figure}

The lexicon maps verbs and noun phrases to grammar rules and to complex and simple concepts respectively. Fig.~\ref{fig:lexicon} shows an illustrating example with a lexical entry on the left and the corresponding grammar unit on the right.  During generation/parsing, the semantic literals listed in the lexicon (here, \sem{Use} and   \sem{useArg2inv}) are used to instantiate the variables (here, \sem{A2} and \sem{Rel}) in the semantic schema (here, \sem{L$_0$:subset(X,L$_1$)   L$_2$:exists(A2,L$_3$) L$_3$:Rel(Y)}). Similarly, the Anchor value (\nl{used}) is used to label the terminal node marked with the anchor sign ($\diamond$) and each coanchor is used to label the terminal node with corresponding name. Thus, the strings \nl{shall} and \nl{be} will be used to label the terminal nodes $V1$ and $V2$ respectively.
Importantly, this separation between grammar and lexicon supports
modularity in that e.g., different lexicons and/or grammars could be
plugged into the system. For the work presented
here, we built this lexicon by applying
regular expressions and a customised NP chunker (the
 NLTK regular expression chunker) to extract verbal and
nominal lexical entries from SIDPs.

\begin{figure}[ht]
\vspace{-1.0em}
\begin{center}
\includegraphics[width=0.6\columnwidth]{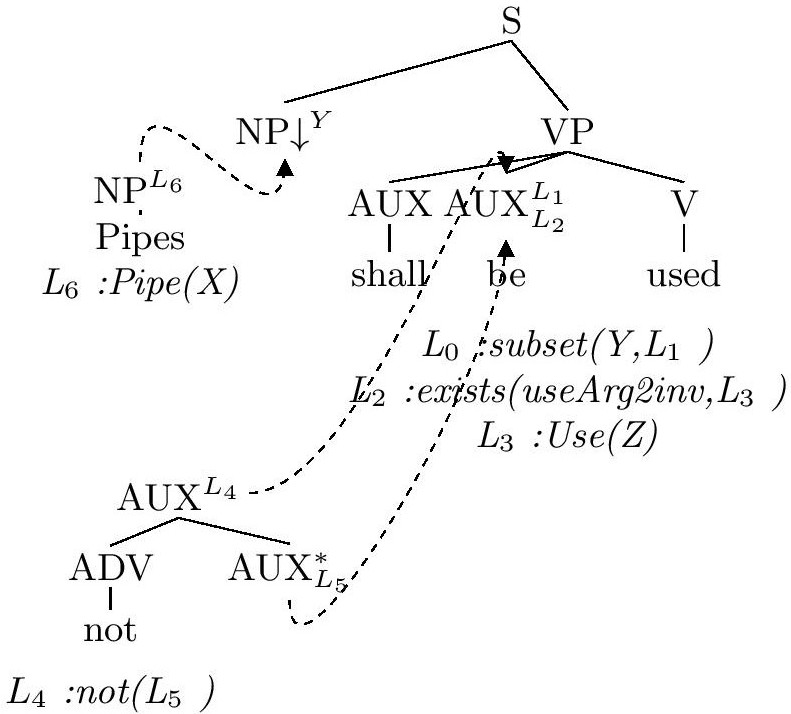}
\vspace{-0.5em}
\caption{\label{fig:examplefbltag} Example FB-LTAG with Unification-Based Semantics. The  variables decorating the tree nodes (e.g., $X$) abbreviate feature structures of the form $[idx:X]$ where $X$ is a unification variable. }
\vspace{-1.0em}
\end{center} 
\end{figure}

The grammar provides a declarative specification of how text relates to meaning (as represented by OWL DL \cite{OWL} formulae).  We use a Feature-Based Lexicalised Tree Adjoining Grammar (FB-LTAG) \cite{gardent2003semantic} augmented with a unification-based flat semantics. Fig.~\ref{fig:examplefbltag} shows an example FB-LTAG for the words ``not", ``pipes" and ``shall be used". An FB-LTAG tree is a set of initial and auxiliary trees which have been lexicalised using the lexicon and can be combined using either substitution or adjunction. Auxiliary trees are trees such as the tree for ``not" which contains a foot node (marked with *) whose category (here AUX) matches that of the root node. Initial trees are trees such as that of ``pipes" and ``shall be used" whose terminal nodes may be substitution nodes (marked with $\downarrow$). Substitution inserts a tree with root category $C$ into a substitution node of the same category. For instance, the tree for ``pipes" may be substituted in the $NP_\downarrow^Y$ node of th ``shall be used" tree. Adjunction inserts an auxiliary tree with foot node category $C$ into a tree at a node of category $C$. For instance, the tree for ``not" may be adjoined into the tree for ``shall be used" at the AUX node. 

\begin{figure}[ht]
\vspace{-1.0em}
\begin{center}
\includegraphics[width=0.6\columnwidth]{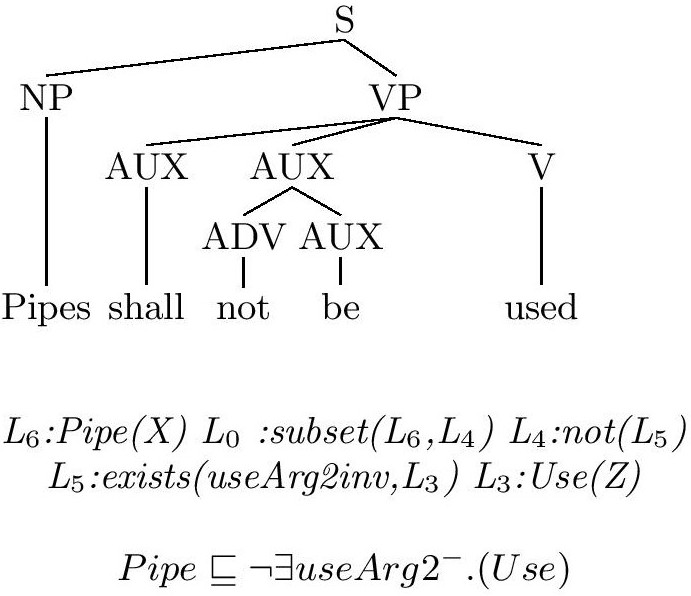}
\vspace{-0.5em}
\caption{\label{fig:derivedtree} Derived Tree. The flat semantics representation produced by the grammar is equivalent to the Description Logic Formula shown.}
\end{center} 
\vspace{-1.0em}
\end{figure}

The parser and the generator exploit the grammar and the lexicon to map natural language to OWL DL formulae (semantic parsing) and OWL DL formulae to natural language (generation) respectively. For instance, given the sentence ``Pipes shall not be used", the parser will first select the grammar trees associated with ``Pipes", ``shall be used" and ``not" and then combines these trees using substitution and adjunction.  As shown in Fig.~\ref{fig:derivedtree}, the semantics derived for the input sentence is then the union of the semantics of these trees modulo unification. Conversely, given the flat semantics shown in the figure the generator will generate the sentence ``Pipes shall not be used" by first, selecting grammar trees whose semantics subsumes the input and then combining them using substitution and operation. The generated sentences are given by the yield of the derived trees whose root is of category S (sentence) and whose semantics is exactly the input semantics. 

While the grammar integrates a so-called flat semantics, as shown in Fig.~\ref{fig:flat2dl}, there is a direct translation from this semantics to OWL functional syntax. Further details about Semantic Parser can be found at:

\begin{center}
\fbox{\bf\small{\url{https://github.com/ModelWriter/Deliverables/tree/master/WP2}}}
\end{center}

\begin{figure*}[ht]
%\addtolength{\belowcaptionskip}{-1mm}
\begin{center}
\small
$$
\tau(\phi)= 
\begin{cases}
    \mbox{ObjectSomeValuesFrom}(\mbox{:R }\tau(C)) & \text{if } \phi = l_i :exists(R,l_j)\;\; l_j :C \\
    \mbox{SubClassOf}(\tau(C_1 ) \; \tau(C_2 )) & \text{if } \phi = l_i :subset(l_j ,l_k )\;\; l_j :C_1 \;\; l_k :C_2 \\
    \mbox{ObjectIntersectionOf}(\tau(C_1 ) \; \tau(C_2 )) & \text{if } \phi = l_i :and(l_j ,l_k )\;\; l_j :C_1 \;\; l_k :C_2 \\
    (\tau(C1) \sqcap \tau(C2)) & \text{if } \phi = l_i :and(l_j , l_k ) \;\; l_j:C1 \;\; l_k:C2\\
    (\tau(C1) \sqcup \tau(C2)) & \text{if } \phi = l_i :or(l_j , l_k ) \;\; l_j:C1 \;\; l_k:C2\\
    \mbox{not}(\tau(C)) & \text{if } \phi = l_i :not(l_j ) \;\; l_j:C\\
    \mbox{R$^-$} & \text{if } \phi = Rinv\\
    \mbox{C} & \text{if } \phi = l_i:C(x)\\
\end{cases}
$$
\normalsize
\vspace{-1.0em}
\caption{\label{fig:flat2dl} Mapping Flat Semantics to Owl Functional Syntax }
\vspace{-1.0em}
\end{center} 
\end{figure*}

\subsection{Inferring Traces using DL Theorem Proving}
\label{subsec:dltp}

We use Hermit theorem prover to detect inconsistencies, entailment and equivalence between two SIDPs $s_1$ and $s_2$. Given the DL formulae $\phi_1$ and $\phi_2$ associated by the semantic parsing process to $s_1$ and $s_2$, we determine these relations as follows: (i) if $\phi_1 \sqcap \phi_2$ is not satisfiable, we infer a \textsc{conflicts} trace between $s_1$ and $s_2$, (ii) if $\neg \phi_1 \sqcup \phi_2$ is satisfiable, we infer a \textsc{requires} trace between $s_1$ and $s_2$, and (iii) if $\phi_1 \equiv \phi_2$ is satisfiable, we infer an \textsc{equals} trace between $s_1$ and $s_2$.

\subsection{Formal Semantics of Trace-types} \label{subsec:formalization} 

\T~is the module of \MW~approach for automated reasoning about traces based on configurable trace semantics, recently described in \cite{ferhat2017SAC:PL}. The tool provides an enhanced text editor to allow users to define new trace types in a restricted form of Alloy~\cite{AlloyBook}, i.e., First-Order Relational Logic. 

In the following, we axiomatize trace semantics based on the informal definition explained in Section \ref{sec:usecase} using \textit{First-order Predicate Logic} with the signature:

\small
\begin{gather*}
    \Sigma_T : \{ =, \in \} \cup \Sigma^1_T \cup \Sigma^2_T \\
    \Sigma^1_T : \{ Artifact, Requirement, Specification \}\\
    \Sigma^2_T : \{ requires, refines, contains, equals, conflicts \}
\end{gather*}
\normalsize

$\Sigma^1_T$ is the set of unary predicate symbols and $\Sigma^2_T$ is the set of binary predicate symbols. For simplicity, we assume that the universe only consists of the type, \textit{Artifact} which is partitioned into disjoint subsets of \textit{Requirement} and \textit{Specification}. From now on, $A$ represents the set of \textit{Artifacts}. $=$ and $\in$ symbols are interpreted and represent \textit{equality} and \textit{membership} respectively. In the following several axiom schemas are listed to formalize Traceability Theory, that is used in the SIDP case. 

Reasoning about \textsc{requires} traces is stated as follows:
\small
\begin{gather}
\hspace*{-1em}\vdash \forall a, b, c \in A \mid (a,b) \in \rsquare \wedge\, (b,c) \in \require \to (a,c) \in \require \label{axiom:requires_1}  \\
\hspace*{-1em}\vdash \forall a, b, c \in A \mid (a,b) \in \require \wedge\, (b,c) \in \rsquare \to (a,c) \in \require \label{axiom:requires_2} \\
\notag\mbox{\quad where \:} \, \square \in \{ \requires, \refines , \contains \}
\end{gather}
\normalsize

{The following axiom schema is being used for generating \textsc{conflicts} traces.}
\small
\begin{gather}
\hspace*{-1em}\vdash \forall a, b, c \in A \mid (a,b) \in \square \wedge (b,c) \in \triangle \to (a,c) \in \triangle \label{axiom:conflict_generation}\\
\hspace*{-1em}\vdash \forall a \in A \mid (a,a) \in \triangle \\
\notag\mbox{\quad where \:} \square \in \{ \requires , \refines , \contains \} \mbox{ and \:} \triangle = \conflicts
\end{gather}
\normalsize

{Reasoning about \textsc{equals} traces:}
\small
\begin{gather}\label{axioms_equals} 
    \vdash \forall a, b, c \in A \mid (a,b) \in \equals \wedge\, (b,c) \in \square \to (a,c) \in \square \\
    \vdash \forall a, b, c \in A \mid (a,b) \in \equals \wedge\, (c,b) \in \square \to (c,a) \in \square \\
    \vdash \forall a \in A \mid (a,a) \in \equals \\
    \notag\mbox{where \:}  \square \in \{ \contains , \requires , \refines , \conflicts \}
\end{gather}
\normalsize

In the following axiom schema, transitivity \eqref{transitivity} is used for reasoning new traces, whereas anti-symmetry \eqref{anti-symmetry} and irreflexivity \eqref{irreflexivity} are used to check consistency.
\small
\begin{gather}
\vdash \forall a, b, c \in A \mid (a,b) \in \square \wedge (b,c) \in \square \to (a,c) \in \square  , \label{transitivity}  \\
\vdash \forall a, b \in A \mid (a,b) \in \square \wedge (b,a) \in \square \to a = b , \label{anti-symmetry}  \\
\vdash \forall a \in A \mid (a,a) \notin \square,  \label{irreflexivity} \\
\notag\mbox{where \:}  \square \in \{ \contains , \requires , \refines \}
\end{gather}
\normalsize

\textsc{contains} traces is left-unique (injective relation) in some scenarios that induces an inconsistency when transitivity axiom \eqref{transitivity} for \textsc{contains} is instantiated in the specification.
\small
\begin{gather}
\vdash \forall a, a', b \in A \mid (a,b) \in \square \wedge (a',b) \in \square \to a = a' \\
\notag\mbox{\quad where \:} \, \square = \contains
\end{gather}
\normalsize

\begin{figure}[ht]
\centering
%\vspace*{-0.5em}
\includegraphics[width=\columnwidth]{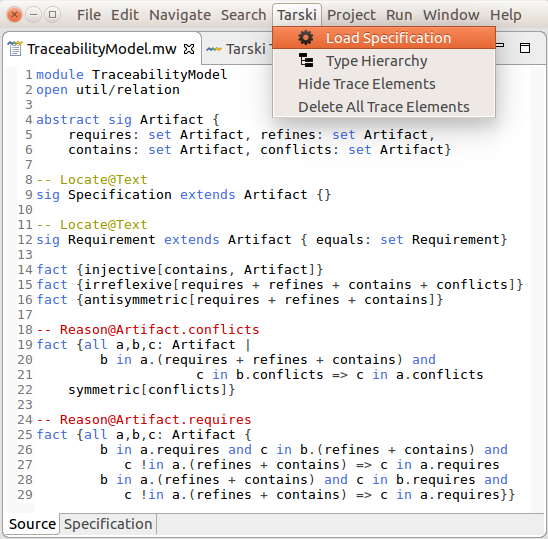}
\vspace*{-1.5em}
\caption{Some Example Trace Types and Trace Semantics in \T}
\label{step:FormalSpecification}
\vspace*{-1.0em}
\end{figure}

We encode above axioms in First-order Relational Logic using the \T's text editor to configure the \T~module (see Figure~\ref{step:FormalSpecification}).

% \begin{comment}
% \end{comment}

\subsection{Inferring Trace Links using Model Finding} \label{subsec:inferrence_traces}
\begin{figure}[ht]
%\vspace*{-1.0em}
\includegraphics[width=\columnwidth]{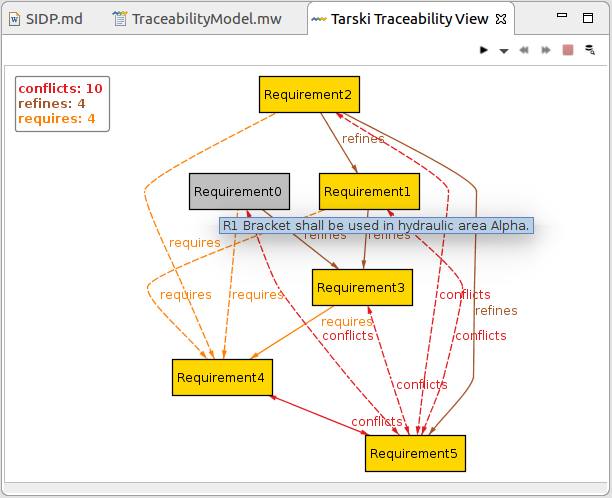}
\vspace*{-1.0em}
\caption{Inferred Relations based on the current snapshot} \label{step:inference}
\vspace*{-1.2em}
\end{figure}

We employ Kodkod~\cite{Torlak:2007, KodKodPhdThesis}, an efficient SAT-based constraint solver for FOL with relational algebra and partial models, for automated trace reasoning using the trace semantics that user provides. Once the user performs reasoning operations about traces, the result is reported back to the user by dashed traces as shown in Fig. \ref{step:inference}. If there exists different solutions, the user can traverse them back and forth. He can also accept the inferred traces, and perform another analysis operation including inferred traces. Further details about Tarski can be found at:

\begin{center}
\fbox{\bf\small{\url{https://modelwriter.github.io/Tarski/}}}
\end{center}
\vspace{-0.6em}

% !TEX root =  Main.tex
\section{Evaluation} 
\label{sec:evaluation}

We evaluate Semantic Parsing approach of \MW~on a dataset of 960 SIDPs provided by Airbus which demonstrate (i) that the approach is robust (97.50\% of the SIDPs can be parsed) and (ii) that DL axioms assigned to full parses are very likely to be correct in 96\% of the cases. Regarding inference phase, since we observed that DL-based reasoning is relatively faster than the SAT-based reasoning in the context of SIDP case, we only focus on the \T~module to evaluate the performance of \MW~approach. Table~\ref{tab:evaluation} shows the solving results of three configurations of the formal trace specification running with Alloy Analyzer \cite{AlloyBook}, KodKod\cite{KodKodPhdThesis} and Z3\cite{z3}. Minisat\cite{Minisat} SAT Solver is chosen for both Alloy (alloy4.2-2015-02-22.jar) and KodKod (Kodkod 2.1) solvers. From Alloy to SMT solver translation for these cases, we employ the translation method proposed by El Ghazi et.al. \cite{ElGhazi2011} and the problems are encoded in SMT-LIB \cite{smtlib} syntax which is fed into Z3 solver. Transitive closure and integer arithmetic are not used in these use cases to fairly benchmark the results with the SMT solver. In SMT-LIB, the logic is set for Equality Logic with Uninterpreted Functions (UF). 

\begin{table}[H] 
\vspace*{-1.0em}
\caption{Comparisons of Several Use Cases for Trace Inferring} \label{tab:evaluation}
\vspace*{-1.0em}
 \begin{center}
   \begin{tabular}{l c c c  r r r}
    \toprule
     & Artifacts & Traces & Inferred &Alloy &KodKod &Z3\\
    \midrule
    $\#1$ &123 &102  &89  &67922 &25668 &40900\\   
    $\#2$ &56  &27   &25  &4428  &84    &480\\
    $\#3$ &42  &103  &75  &724   &1     &1460\\
    \bottomrule
   \end{tabular}
  \end{center}
 \vspace*{-1.0em}
\end{table}

Evaluation results are obtained on a machine, that runs 64 bit debian linux operating system with 8 GB of memory and 2.90GHz Intel i7-3520M CPU. Solving times are indicated in milliseconds. The best results are obtained by the direct use of KodKod API since to find satisfiable models, KodKod allows us to configure lower and upper bounds for the solution space employing different pre-processing techniques such as slicing, incremental upper bounds and unrolling transitive closures. The evaluation shows that our tool is practical and beneficial in industrial settings to specify trace semantics for automated trace reasoning. We plan to conduct more case studies to better evaluate the practical utility and usability of the platform. 

% !TEX root =  Main.tex
\section{Related Work} 
\label{sec:relatedwork}

Many existing works on semantic parsing describe the task of obtaining axiomatic representation of natural language sentences. However, they suffer from two main limitations: (i) use of controlled languages such as Attempto Controlled English\cite{kaljurand2007verbalizing} (e.g. \cite{tablan2006user,bernstein2005querying}) and/or (ii) inability to deduce complex axioms involving logical connectives, role restrictions and other expressive features of OWL (e.g. \cite{buitelaar2005ontology,ruiz2005automatic}), as noted in \cite{volker2007acquisition}. In contrast, we work on human authored real-world text (Airbus SIDPs) and produce complex OWL axioms involving the following DL constructs: $\top$ (the most general concept), disjunction, conjunction, negation, role inverse, universal and existential restrictions. Moreover, we extended the scope of our application by deducing traces among the semantic parse outputs. Such traces were then used as baseline input to \T~platform which could infer additional traces propagating over the whole system.

Similarly, several approaches and tools have been proposed for automated trace reasoning using the trace semantics~\cite{TraceabilityOperationalSemantics2005, EgyedG05, Egyed03, Cleland-HuangCC03, lamb2011, SemanticTrace2011, FormalTraceSemantics2014, Drivalos2008}. These approaches employ a predefined set of trace types and their corresponding semantics. For instance, Goknil et al.~\cite{SemanticTrace2011} provide a tool for inferencing and consistency checking of traces between requirements using a set of trace types and their formal semantics. Similarly, Egyed and Gr{\"{u}}nbacher~\cite{EgyedG05} propose a trace generation approach. They do not allow the user to introduce new trace types and their semantics for automated reasoning. In the development of complex systems, it is required to enable the adoption of various trace types, and herewith automated reasoning using their semantics. \T~module of \MW~allows the user to interactively define new trace types with their semantics to be used in automated reasoning about traces.

% !TEX root =  Main.tex
\section{Conclusion} 
\label{sec:conclusion}

We presented an integrated platform for automatically mapping natural language text to trace types and performing further inferencing on those traces. Starting with the semantic parser module, we showed how complex axioms could be derived to represent text coming from real world use case. We identified the traces among the parse outputs and fed it to the \T ~tool. The \T ~tool, in turn, allowed users to specify configurable trace semantics for various forms of automated trace reasoning such as inferencing and consistency checking. The key characteristics of our tool are (1) automatic identification of traces existing in texts using semantic parsing (2) allowing user to define new trace types and their semantics which can be later configured, (3) deducing new traces based on the traces which the user has already specified, and (4) identifying traces whose existence causes a contradiction.

% !TEX root =  Main.tex

\section*{Acknowledgment}
This work is conducted within ModelWriter project\cite{ModelWriter} labeled by the European Union's EUREKA Cluster programme ITEA and partially supported by the Scientific and Technological Research Council of Turkey under project \#9140014, \#9150181 and Industry and Digital Affairs of France, Directorate-General for Enterprise under contract \#142930204. The authors would like to acknowledge networking support by European Cooperation in Science and Technology Action IC1404 "Multi-Paradigm Modelling for Cyber-Physical Systems".

%\newpage


%\bibliographystyle{IEEEtran}
\bibliographystyle{IEEEtran}
\bibliography{IEEEabrv,IEEEexample}

\begin{comment}
    the envisioned users;
    the software engineering challenge it proposes to address;
    the methodology it implies for its users; and
    the results of validation studies already conducted for mature tools, or the design of planned studies for early prototypes.
\end{comment}

%\newpage
% !TEX root =  Main.tex
\appendices

% This next section command marks the start of
%Appendix A, and does not continue the present hierarchy
\section{Availability \& Open Source License} \label{apx:availability} 

\noindent This work is being developed under technical Work Package 2 (Semantic Parser - Text Part), Work Package 3 (Tarski - Model Part), and Work Package 4 (Federated Knowledge Base) within ModelWriter project, labeled by the European Union's EUREKA Cluster programme ITEA (Information Technology for European Advancement). Further details about the project can be found at: 

\begin{center}
\fbox{\bf\small{\url{https://itea3.org/project/modelwriter.html}}}
\end{center}

\noindent A video demonstration of \MW~is available shows the use of \MW~in the context of the industrial use case \E~presented in the paper. The video is available at:
\begin{center}
\fbox{\bf\small{\url{https://youtu.be/TcVCosW8HkU}}}
\end{center}

\noindent The source codes files and datasets of \MW~are publicly available for download and use at the project repository. \T~and SemanticParser are components of \MW~platform. Source codes, screencast and datasets regarding the project are also available and can be found at: 

\begin{center}
\fbox{\bf\small{\url{https://github.com/ModelWriter/Deliverables/tree/master/WP2}}}
\end{center}

\begin{center}
\fbox{\bf\small{\url{https://modelwriter.github.io/Tarski/}}}
\end{center}

\begin{center}
\fbox{\bf\small{\url{https://github.com/ModelWriter/Source/}}}
\end{center}

\begin{center}
\fbox{\bf\small{\url{https://github.com/ModelWriter/Demonstration/}}}
\end{center}

\vspace*{0.1cm}
\noindent \MW~is distributed with an open source software license, namely \textit{Eclipse Public License v1}. This commercially friendly copyleft license provides the ability to commercially license binaries; a modern royalty-free patent license grant; and the ability for linked works to use other licenses, including commercial ones.

\begin{center}
\fbox{\bf\small{\url{https://github.com/ModelWriter/Source/blob/master/LICENSE}}}
\end{center}

% Appendix B
\section{Tool Demonstration Plan} \label{apx:demonstration} 
There will be four parts to our presentation: (1) motivation and industrial use cases, (2) overview of the approach and tool architecture, (3) demonstration walktrough, and (4) evaluation.  Parts 1, 2 and 4 are presented using slides while Part 3 is presented as a demo using the industrial use case scenario described in Section \ref{sec:usecase}. To present these parts, we use a combination of slides, animations, and a live demo. In the following subsections, we provide further details about our presentation plan. 

\subsection{Motivation \& Challenges} \label{Motivation}

\subsubsection{Motivation} 
We will emphasize the importance of traceability by introducing "\textit{DO-178C Software Considerations in Airborne Systems and Equipment Certification}"~\cite{DO178C} from aviation industry.

\subsubsection{Industrial Use Cases} 
We will briefly describe the challenges of \textit{Traceability Analysis Activities} faced in industry by introducing industrial use cases from Airbus. We will explain the importance of \textit{semantically meaningful traceability}, \textit{traceability configuration} and \textit{automated traceability analysis} in industry.

\subsection{Tool Overview} \label{Overview}

\subsubsection{Overview of the Solution.} 
We will explain the approach and the user workflow of \MW~by following the steps of Section \ref{sec:approach}.

\subsubsection{\MW~Features} 
We will briefly explain tool features such as semantic parser and its reasoning engine and configurable automated traceability analysis using animated slides by giving concrete examples from the industrial use case, namely \case~presented in the paper.

\subsection{Walk-trough of the Tool Demonstration} \label{Demonstration}

In this section, we will perform a live demonstration aligns with the industrial use case \E, which is illustrated in Section \ref{sec:usecase}.

\subsection{Evaluation and Lessons Learned} \label{Evaluation}

We conclude with a summary that presents the evaluation results and the lessons learned.

\end{document}